# Finnish primary school students' conceptions of machine learning

Pekka Mertala[1], Janne Fagerlund[1], Jukka Lehtoranta[2], Emilia Mattila, Tiina Korhonen[3]

[1]University of Jyväskylä, Department of Teacher Education; [2]No affiliation; [3]University of Helsinki, Faculty of Educational Sciences

Corresponding author
Pekka Mertala
Department of Teacher Education, University of Jyväskylä, Jyväskylä, Finland
Alvar Aallon katu 9, 40014, Jyväskylä, Finland
pekka.o.mertala@jyu.fi

## Abstract

**Background and context**

Language is a powerful tool in shaping thought about abstract domains, including the realm of computing and digital technologies. Review of research on children's machine learning (ML) conceptions suggests that participating children were often provided with conceptual clues about the principles of ML before being asked about their conceptions (i.e., asking them how one could teach a computer). Since the term ML is not explicit about who is learning and from whom, this procedure has arguably steered their answers: with less nudging instruction, ML could also be understood as a process in which a human uses a machine, a computer, for instance, for learning purposes.

**Objective**

This study investigates what kind of conceptions primary school students have about ML if they are not conceptually "primed" with the idea that in ML, humans teach computers.

**Method**

Qualitative survey responses from 197 Finnish primary schoolers were analyzed via an abductive method.

**Findings**

We identified three partly overlapping ML conception categories, starting from the most accurate one: ML is about teaching machines (34%), ML is about coding (7.6%), and ML is about learning via or about machines (37.1%).

**Implications**

The findings suggest that without conceptual clues, children's conceptions of ML are varied and may include misconceptions such as ML is about learning via or about machines. The findings underline the importance of clear and systematic use of key concepts in computer science education. Besides researchers, this study offers insights for teachers, teacher educators, curriculum developers, and policymakers.

## Introduction

This qualitative study addresses the research question: What kinds of conceptions do primary school students have about machine learning? Conceptions are understood as the ideas individuals employ to make sense of the world around them (Marton, 1981; Thompson & Logue, 2006). These are essentially explanations and hypotheses concerning the essence, origins, and functioning of various phenomena. Scientific conceptions, often considered correct explanations, provide a factual, science-based understanding of how and why things operate and what they are (Vygotsky, 1987; Edwards et al., 2018).

Given our highly digitalized lifeworld, children's conceptions of digital technologies—including computers, coding, the Internet, and search engines—have frequently been explored in research (e.g., Babari et al., 2023; Kodama, 2016; Edwards et al., 2018; Eskelä-Haapanen & Kiili, 2019; Mertala, 2019, 2020; Rubegni et al., 2022; Rucker & Pinkwart, 2016; Wennås Brante & Walldén, 2023). The 2020s have seen an intensive focus on studying children's perceptions of Artificial Intelligence (AI) (e.g., Kim et al., 2023; Kreinsen & Schultz, 2021; Marx et al., 2023; Mertala & Fagerlund, 2024; Mertala et al., 2022; Ottenbreit-Leftwich et al., 2021, 2022; Oyedoyin et al., 2024; Solyst et al., 2023; Vandenberg & Mott, 2023).

While the current era is often characterized by the prominence of machine learning (ML; (e.g., Audry, 2021; Ourmazd, 2020; Valtonen et al., 2019)—a key methodological and technological foundation of AI—research focusing on children's conceptions of ML is comparatively rare (cf. Druga & Ko, 2021; Hitron et al., 2018; Muhling & Große-Bölting, 2023; Vartiainen et al., 2021). Research has predominantly aimed at determining how and what children should be taught about ML, to enhance their comprehension of the contemporary digitalized world and their ability to function within it (e.g., Sanusi et al., 2023; Shamir & Evin, 2022; Vartiainen et al., 2021). Even though such efforts are crucial for developing research-based ML curricula, from a constructivist viewpoint, understanding students' pre-existing conceptions is vital: it informs us about potential misconceptions that could hinder the learning of new concepts, leading to further misunderstandings (Biber et al., 2013). This is particularly relevant for AI and ML, where misconceptions may impair the effective use of these technologies, for example, in prompt engineering.

Furthermore, a review of prior research on children's ML conceptions (e.g., Druga & Ko, 2021; Hitron et al., 2018; Marx et al., 2023; Muhling & Große-Bölting, 2023; Vartiainen et al., 2021) reveals that participants often received conceptual clues about machine learning principles before expressing their conceptions. This approach may have inadvertently guided their responses. For instance, students in the study by Vartiainen et al. (2021) were instructed to "draw and/or write... thoughts and ideas about *how one could teach a computer* (italics original)," with the instructions clearly implying that computers/machines can learn and be taught. Similarly, Muhling and Große-Bölting (2023) asked participants to "describe how you imagine *machines to learn to play* X and O, (italics added)" providing explicit cues about the learning capabilities of machines.

While placing the focus on individual words may seem trivial, it is important to acknowledge that "language is a powerful tool in shaping thought about abstract domains" (Boroditsky, 2001, p. 1). For instance, a recent study (Mertala & Fagerlund, 2024) on Finnish primary school students' misconceptions of AI found that some students conceptualized AI as a kind of cognitive process or an act or an action that people engage in. The authors reasoned that language most likely has shaped these conceptions:

> In the Finnish language, AI is called "tekoäly," a compound word that unites the terms "teko" and "äly." While the latter term, "äly," translates as intelligence, "teko" is a more ambiguous concept: besides "artificial," the word also refers to "an act" and "an action." Indeed, many of the non-technological misconceptions described different kinds of cognitive acts and actions (e.g., regulation of immediate and intuitive instincts). (Mertala & Fagerlund, 2024, p. 6.)

## 2. Conceptual Notions on Machine(s and) Learning

Machine and learning. Among these two concepts, "machine" is relatively straightforward. The Merriam-Webster dictionary describes a machine as "a mechanically, electrically, or electronically operated device for performing a task." For instance, a washing machine is specifically designed to wash clothes and textiles. However, this term's complexity increases when combined with others. To illustrate, consider the concept of AI.

According to Merriam-Webster, "artificial" refers to something humanly contrived, often based on a natural model, thus standing in contrast to what is natural. An artificial flavor, for example, may mimic strawberries without containing any real strawberries, achieved by creating a specific chemical compound ($C_{12}H_{14}O_3$) in a laboratory. However, while an artificial strawberry flavor can be defined by a chemical formula, intelligence lacks a universal definition. Although intelligence has a fairly concrete meaning colloquially, its definition becomes more complex in scholarly discourse (Legg & Hutter, 2007a, 2007b).

Gardner's theory of multiple intelligences serves as a notable example. Gardner posits the existence of seven or eight distinct intelligences, depending on the publication: linguistic, logical-mathematical, spatial, musical, naturalist, bodily-kinesthetic, interpersonal, and intrapersonal—the latter being a more recent addition (Davus et al., 2011; Gardner & Hatch, 1989. Davis et al. (2011) describe intrapersonal intelligence as the "ability to recognize and understand one's own moods, desires, motivations, and intentions". This is something AI is not capable of. While personal assistants like Siri can provide witty responses such as "I'm not sure what you have heard but virtual assistants have feelings too" (Taubenfeld, 2023), they do not possess emotions, moods, or equivalent.

Similar to intelligence, "learning" can refer to a broad array of phenomena (Biesta, 2015), with definitions varying significantly across and within disciplines, and new definitions continue to be proposed (e.g., Barron et al., 2015; De Houwer et al., 2013; Lachmann, 1997). Individuals often hold colloquial conceptions of learning that diverge from scholarly definitions. Children, for instance, conceptualize learning as the acquisition of skills, knowledge, or understanding (Mertala, 2022; Pramling, 1998; Sandberg et al., 2017). An interesting consideration is how these varying conceptions of learning align with the principles and current boundaries of machine learning. For example, conversational large language models (LLMs), like ChatGPT, appear capable of understanding or knowing things, albeit within the confines of their programming to respond based on their training data and feedback from human trainers.

As detailed on OpenAI's website, ChatGPT is optimized for dialogue through Reinforcement Learning with Human Feedback (RLHF), a method that guides the model towards desired behavior using human demonstrations and preference comparisons. However, this does not imply that ChatGPT understands the reasons behind its responses or grasps the content or meaning of the text. It learns to predict the next word, sentence, or paragraph based on its training data and feedback from human trainers.

Beyond reinforcement learning, machine learning encompasses various methods such as supervised, unsupervised, semi-supervised learning, and neural networks, each with distinct principles and applications. Reinforcement learning with human feedback is merely one branch of the broader field of reinforcement learning. Table 1 summarizes the key principles of common ML techniques, drawing on the reviews by Mahesh (2022) and Sarker (2021).

Table 1. The key-principles of common ML techniques (Mahesh, 2022; Sarker, 2021)

| ML technique | Key principles |
| --- | --- |
| **Supervised learning** | Supervised learning is a machine learning approach where a model learns to map inputs to outputs by studying labeled training data. It relies on external guidance and involves splitting the dataset into training and test subsets. The training data contains examples with known outcomes, and various algorithms learn patterns from this data to make predictions or classifications on the test data. Examples of supervised learning techniques are decision trees and naive Bayes. |
| **Unsupervised learning** | Unsupervised learning analyzes unlabeled datasets without the need for human interference, i.e., a data-driven process. This is widely used for extracting generative features, identifying meaningful trends and structures, groupings in results, and exploratory purposes. The most common unsupervised learning tasks are clustering, density estimation, feature learning, dimensionality reduction, finding association rules, anomaly detection, etc. Examples of unsupervised learning applications are Principal component analysis and K-means clustering. |
| **Semi-supervised learning** | Semi-supervised machine learning is a combination of supervised and unsupervised machine learning methods as it operates on both labeled and unlabeled data. The ultimate goal of a semi-supervised learning model is to provide a better outcome for prediction than that produced using the labeled data alone from the model. Some application areas where semi-supervised learning is used include machine translation, fraud detection, labeling data and text classification. |
| **Reinforcement learning** | Reinforcement learning is an area of machine learning concerned with how software agents ought to take actions in an environment in order to maximize some notion of cumulative reward. This type of learning is based on reward or penalty, and its ultimate goal is to use insights obtained from environmental activists to take action to increase the reward or minimize the risk. It is used for training AI models that can help increase automation or optimize the operational efficiency of sophisticated systems such as robotics, autonomous driving tasks, manufacturing and supply chain logistics. |
| **Neural networks** | A neural network is a series of algorithms that endeavors to recognize underlying relationships in a set of data through a process that mimics the way the human brain operates. In this sense, neural networks refer to systems of neurons, either organic or artificial in nature. Neural networks can adapt to changing input; so the network generates the best possible result without needing to redesign the output criteria |

Of course, we are not suggesting that the table above should be included in the K-12 curriculum as such. However, being aware of the principles of ML even at a rudimentary level is important. The models underpinning current popular applications that utilize ML, such as ChatGPT and Midjourney, are black boxes, that is, systems that are only observed in terms of their input and output, hiding the internal mechanisms. This is problematic for developing a clear understanding of the inner workings of AI and ML, and ways in which input controls, such as natural language prompts, are transformed into output, become unclear, weakening the ability to make effective use of the tool. For example, when using ChatGPT to make a query, if one has an anthropomorphic conception of AI (see Mertala & Fagerlund, 2024), one may bring the communication practices of natural languages and the more common conventions of human interaction into the 'conversation' with the user interface (i.e., prompting; see also Jochmann-Mannak et al., 2010; Kammerer & Bohnacker, 2012). But the language model does not work in the same way: it requires specific kinds of technical practices that specifically support its effective use. In other words, unstructured prompts stemming from misunderstanding the technology may result in inaccurate responses (possibly 'hallucination' from the LLM), which, in turn, may undesirably result in the formation of false beliefs about the topic of the query.

## 3. The present study

### 3.1. Context and participants

This study was conducted within the context of an AI pilot project (2022–2024) organized by Innokas, a Finnish national school innovation network for teachers, coordinated by the University of Helsinki, and funded by the Finnish National Agency for Education. The participants comprised Finnish primary school students (N=197), with data collected during the project's first year. At this time, teachers had primarily received professional AI training and had executed small pilot projects within their educational contexts. In the second year, they proceeded to plan and implement more structured AI-themed interventions in their classrooms. Consent for student participation was obtained from their legal guardians, adhering to guidelines set by the Finnish National Board of Research Integrity (2019). Table 2 outlines the distribution of participants across grades and genders.

Table 2: Participants

| Grade | Girls | Boys | Not reported | Total |
|---|---|---|---|---|
| 4 | 18 | 29 |   | 47 |
| 5 | 14 | 22 |   | 36 |
| 6 | 16 | 24 | 1 | 41 |
| 7 | 12 | 5 |   | 17 |
| 8 | 13 | 32 | 1 | 46 |
| 9 | 4 | 6 |   | 10 |

| **Total** | 77 | 118 | 2 | 197 |

19 participants were Swedish speakers, 10 participants responded in English and the remaining 168 responses were in Finnish.

### *3.2. Data and analysis*

The data were collected via an online survey approximately at the midpoint of the project. The participating students had therefore taken part in little to no formal AI education on behalf of the project at the time of data collection. The survey contained both open and closed questions. The data for the present paper contain students' responses to a prompt: "write, in your own words and as versatile as possible, what the following words mean: artificial intelligence, machine learning, and data." Explanations for each term were written in separate text boxes (see Table 3). The data were analyzed via abductive content analysis, which combines deductive and inductive reasoning (Grönfors, 2011) through constant comparison (Boeije, 2002). In this study, constant comparison included the following intertwined phases:

**Comparison between theory and data**: In abductive analysis, the role of deduction is to offer theoretical threads, which are complemented and/or refined via interpretive inductive analysis (Mertala, 2020). The main theoretical threads were the different forms of machine learning (Mahesh, 2022) and the premise that machine learning can also be understood as learning via or about machines (Mertala & Fagerlund, 2024). For example, Student 97's (rivi boy 7th grade) response "learning with the help of a machine, for example, working with a computer," was coded as ML being about learning with machines via theory-driven reading.

**Comparison within the data of an individual participant** refers to the close reading of the data provided by each student. During this phase, we noticed that an individual student could express possessing conceptions from more than one category. Student 19 (boy 8th grade), for instance, wrote that "well you learn new things with the help of the machine and teach the machine new things," which implies that for him, ML was about both teaching a machine and learning with machines.

While our focus was on machine learning, we included students' responses to the AI and data prompts in the analysis as well. Students' responses to the ML prompt served as primary data, which was complemented by their responses to AI and data prompts (secondary data). The rationale behind this choice was that the concepts are interlinked: machine learning is a subset of AI, and ML always requires data (be it labeled or unlabeled). For example, if a student wrote that ML means that machines learn from data, examples of what the student understands by data were sought from the responses to the other prompts as well. That said, due to the mediating role of language (Boroditsky, 2001; Mertala & Fagerlund, 2024), students' responses to the ML prompt were always the primary target of the analysis. Let us use the two excerpts presented in Table 3 as an example.

Table 3: Data examples

|  | **Artificial intelligence** | **Machine learning** | **Data** |
|---|---|---|---|
| Student 98 (girl 7th grade) | [Artificial intelligence] is, for example, a computer's or a phone's "intelligence", which can do various things it has been taught better, more accurately and faster than a human. | For example, learning via a computer | knowledge contained and used on the internet. |
| Student 45 (boy 6yh grade) | Artificial intelligence is a robot that has been programmed by humans and by time it analyzes data to give accurate answers to what we ask of it. There are some examples of how we use AI in daily life like Facial recognition, ChatGPT even when we like videos on Youtube it registers that data to find more videos like it. | Machine learning can be used to analyze data and learn from its mistakes and can be used to develop better results. | meaning information also like how a robot stores data that it has learn from humans. |

Some narrative soothing (Polkinghorne, 1995), such as correcting misspelled words, is done to improve the narrative flow of the data extracts.

Student 98's conceptualization of AI includes a description that AI can perform tasks it has been taught, suggesting an understanding of machines as learnable/teachable entities. However, her characterization of ML as "learning via a computer" led to coding the response as "ML is about learning with machines." Such implicit conceptions are further addressed in the findings section (see section 4.1).

Regarding student 45, he articulates a view that ML enables AI-powered machines and solutions that have data stored in them. His choice of words, "learn from its [data's] mistakes," suggests that he conceptualizes ML as learnable machines (ML is about teaching machines). Furthermore, his treatment of data as AI's knowledge base ("robot stores the data it has learned from humans" in the column: Data) echoes a previously identified misconception that AI's "knowledge"—or that of computers in general—is preinstalled in the machine (Mertala & Fagerlund, 2024; Rucker & Pinkwart, 2016). This discussion also serves as a **comparison between data and theory**, where data-driven interpretations are juxtaposed with additional research literature.

Lastly, **comparison between the data from different participants** was conducted to identify the distribution of different conceptions within the sample. This was done by counting the frequencies of the different conceptions. 154 of the students (78.2%) provided an explanation of ML, whereas the remaining 43 students (21.8%) responded with "I don't know" or equivalent. The frequency of different conceptions is provided in context in the following section.

## 4. Findings and Discussion

The findings of this study are presented in three subsections introduced in Figure 1. We start from the most accurate conceptions (ML is about teaching machines) and continue to the less accurate ones (ML is about coding, ML is about learning via or about machines). Discussion with relevant empirical and theoretical literature is embedded within the findings.

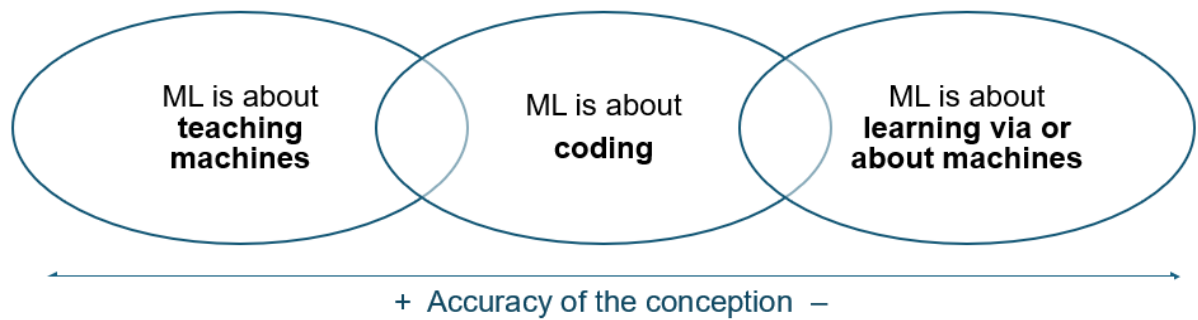

Figure 1: Summary of students' conceptions of machine learning

### *4.1. ML is about teaching machines*

Altogether, 67 students (34.0%) expressed conceptions of ML being about teaching machines, which can, in principle, be considered the most factual of the three categories. The descriptions, however, were far from identical as there was a notable variety with regard to their details, highlighting a more nuanced granularity in exactly how accurate the students were able to provide. With many students, the responses stayed on a general level and included examples such as:

> "You teach some machine" (Student 36; girl, grade 4)

> "A machine that is capable of learning different things" (Student 190; boy, grade 8)

> "Machine learning is when a machine is made to, for example, recognize two colors from each other" (Student 40; boy, grade 4)

In each of the examples, the students explicitly express that in ML the machine is the one that learns or is being taught. That said, none of the excerpts provide a concrete explanation of what the process of ML entails. Even though the last excerpt is the most specific one (as it provides an example of the actual task), it fails to outline what it actually means to make a machine (learn to) identify colors.

Another group of students expressed more accurately that in ML machines learn or are taught with data. Student 14 (boy, grade 8), for instance, wrote that

> "Data is fed into a machine, e.g., pictures of basketballs. It analyzes the images and, using artificial intelligence, learns to identify the basketball."

The choice of words has notable resemblance to supervised learning: the pictures of basketballs serve as labeled training data the algorithm (called AI by the student) uses to identify the characteristics of a basketball. What is missing from the description is the test data (i.e., a set of pictures of different balls and round objects), which is used to evaluate the accuracy of the model, hinting at a slightly deficient yet nevertheless comparably scientific conception of ML.

Student 23 (girl, grade 8) expressed a roughly similar view as she wrote that in ML "a machine learns where a person's eyes or ears are, for example, through selfies. And what kind of features different people have." While she does not use the exact term "data", her response contains an implicit reference that in ML a machine uses data, namely selfies, in order to learn to recognize the core characteristics of a human face.

Other students mentioned that data plays an integral role in teaching machines but were either unable to explain how the data is used or provided inaccurate or incorrect statements. The following excerpt from student 50 (girl, grade 6) is an illustrative example of an inability to explain how the data is used: "I don't know, but my guess is that it means machines' learning from the data given to them." The following excerpt from student 55 (girl, grade 6), in turn, is an example of an incorrect statement. According to her,

> "Machine learning is when a human puts data into the bot. Humans put like one information about something and the bot only does or says what the human put into it. For example, if a scientist puts some data into a bot about math it will know stuff about math."

Even though the answer reflected the categorically most factual ML conception, her choice of words suggests that the ML algorithm is only capable of repeating or retaining the information that people have installed in it.

For a clearer comparison, student 50 expressed that ML learns from the data, whereas student 55 seems to think that via ML, the bot learns the data it is provided. An LLM-powered chatbot provides a useful example to address this difference in a more concrete way. Simply put, an LLM is fed a massive amount of textual data from which it learns the patterns and structures of natural language. As a result, the chatbot can compose new text. It does not reproduce the source texts as such as suggested by student 55.

Conceptualizing digital technologies as omniscient databases with pre-installed knowledge is a common finding in research exploring children's (Kim et al., 2023; Mertala & Fagerlund, 2024; Muhling & Große-Bölting, 2023) and adults' (Selwyn & Gallo-Cordoba, 2022) conceptions of AI and computers in general (Rucker & Pinkwart, 2016). Lastly, three students (1.5%) described the functions of AI in a way that showcased an understanding of the rudimentary principles of ML. However, they were not familiar with the concept of ML: their conceptualizations of ML were either about learning with or about machines (see Table 3) or they outrightly stated that "I don't know" (what ML is) (Student 26; girl, grade 8).

### *4.2. ML is about coding*

A total of 15 students (7.6%) expressed conceptions that ML has something to do with coding. As illustrated in Figure 1, these conceptions partly overlapped with other categories. In general, coding plays a notable role in ML: algorithms must have been written by someone (or at least, the person must have prompted a generative AI to produce such code). Indeed, some of the responses included rather explicit references to the role of coding or algorithms in ML applications. Student 35 (boy, grade 4), for instance, argued that ML is "coding, that is, teaching a machine." Put differently, the rationale here is that coding is a way to teach the machine to do something, which is a description of traditional rule-based programming that stands in contrast with data-driven ML methodologies.

Student 57 (girl, grade 6), in turn, wrote in a similar fashion that; "I am not sure, but I would guess [ML means] coding a machine or when you tell something to a machine and it subsequently works in that way." In her response, she parallels coding with telling something to a machine, both of which she recognizes as ways to make machines master certain tasks. Students in Muhling and Große-Bölting's (2023) study expressed roughly similar views when they were asked to explain how machines learn how to play X and O. For instance, one of their participants said that "the game is learned by the machine as code" (Muhling & Große-Bölting, 2023, p. 6), which implies a belief that the programmer has coded the algorithm to conduct certain moves in certain game circumstances.

Other students in our study conceptualized ML as learning to code, which also overlaps with the category "ML is about learning via or about machines" discussed in the following section. The following extract serves as a representative example of the data,

> "Machine learning is learning to use a machine and getting to know machines. Machine learning can include many things from coding to using different applications" (Student 88; boy, grade 8).

One explanation for the emergence of the conception that ML is about coding is the inclusion of (traditional rule-based) programming education in the currently effective Finnish National Core Curriculum for Basic Education (NAE, 2014). Programming was (re)introduced as a cross-curricular theme to be enacted, in principle, within every subject (with emphasis on mathematics and crafts) (Mertala et al., 2020). While the actual implementation of programming education evidently lags behind (Fagerlund et al., 2022; Tanhua-Piiroinen et al., 2020), it is quite possible that the students had been taught the rudiments about computers as programmable machines and the role of code in programming as part of their studies, which can contribute to a notion of traditional rule-based programming being ubiquitously in the technical heart of digital devices.

### *4.3. ML is about learning via or about machines*

The third category was ML as learning via or about machines, mentioned by 73 students (37.1%). As discussed in the previous section, ML as learning to code was one example of learning about machines. As can be seen from the following responses, learning about machines was mainly conceptualized as functional skills that allow one to operate a machine, which most often meant computer use.

> "I think machine learning is how one can use a computer" (Student 72; boy, grade 5)
>
> "How one uses computers" (Student 142; boy, grade 5)
>
> "Learn to use a computer and the web" (Student 95; girl, grade 7)
>
> "Learning to use different kinds of machines" (Student 110; boy, grade 6)

Typical responses about learning via machines included the following:

> "Machine learning is that you learn something from a machine" (Student 65; boy, grade 5)
>
> "That a robot teaches a person" (Student 69; boy, grade 5)
>
> "Learning things with information technology" (Student 137; girl, grade 9)
>
> "That you study with a computer" (Student 149; girl, grade 4)
>
> "For example, in school, you don't study with a book but the materials are on the web" (Student 81; girl, grade 8)

One explanation relates to the Finnish and Swedish languages. In Finnish, the computer is called "tietokone" (literal translation: "knowledge machine"). The latter term "kone" (machine) in the compound word is the same as in "koneoppiminen" (machine learning), providing a linguistic explanation for this misconception. The same applies to Swedish as well, the second official language in Finland and the language of 19 of the respondents (see Table 2). In Swedish, the computer can be called "datamaskin" (literal translation: "data machine"), and machine learning is called "maskinlärning". Similar findings were observed in the study by Mertala and Fagerlund (2024) in terms of AI, where the first term in the Finnish compound word "tekoäly" (AI) can mean both "artificial" and "an act". Similarly, Finnish preschoolers used such conceptual similarities in reasoning what programming is (Mertala, 2019).

The prevalence of such statements, especially the last one ("For example, in school, you don't study with a book but the materials are on the web"), are likely to be bound to the students' experiences and observations in digital education. Supporting children's digital skills is an explicit objective in the Core Curriculum for Basic Education (NAE, 2014), and thus, Finnish schools are rather digitalized in terms of the availability of devices and software: almost all schools have a wireless internet connection, and there is a tablet for every fourth student (Tanhua-Piiroinen et al., 2020). According to teachers, the most common ways to use technology are information search on the internet, the use of pedagogical online materials, and the use of digital learning environments (Tanhua-Piiroinen et al., 2020; see also Oinas et al., 2023). Furthermore, while the use of social robots in schools is not (yet) mainstream, they are covered regularly in media (e.g., Punkari, 2022; Sillanpää, 2018; Rantala, 2019).

## 5. Concluding remarks

### *5.1. Reflection on the findings*

In this study, we explored Finnish primary and secondary school students' conceptions of ML by analyzing qualitative questionnaire data via abductive analysis. We found that students' conceptions varied from rather accurate definitions (ML is about teaching machines via specific data-utilizing techniques) to what would be fair to describe as misconceptions (ML is about learning via or about machines). However, we found granularity among and between the categories, demonstrating that even the more accurate definitions involved misconceptions and altogether painting a picture of a vast variety of different ways in which students may explain and hypothesize what ML is.

While relatively many students associated ML with teachable and learnable machines, a majority described the process of learning or teaching in a rather superficial manner. This lack of detail is a rather understandable (even expectable) finding. ML solutions are often referred to as "black boxes," as their functional principles are not visible to the general public. Put differently, while we use ML-based applications and services on a daily basis, they remain highly opaque as typically only the input and output are visible to the user, which makes it difficult to make sense of 'what happens in the engine room.'

A somewhat more surprising finding, in turn, was the high number of responses that associated ML with learning about or via machines. While the premise that machine learning can initially be understood also as learning via or about machines was one of the theoretical threads of the analysis—and the research process in general—the fact that more than one-third of the participants shared this misconception requires careful and profound reflection.

The prevalence of this misconception in our data suggests that when conceptual clues like "explain how one could teach a computer" (Vartiainen et al., 2021) and "describe how you imagine machines to learn" (Muhling and Große-Bölting, 2023) are not used, children's initial conceptions about ML can differ notably from the scientific concept and be guided by their experiences and observations, like the use of digital technologies for learning purposes in school. Our findings are partially in contrast with Muhling and Große-Bölting's (2023, p. 7) observation that "almost all of the students [in their study] actually had some conception of machine learning that they were able to explicate upon in a given context." We would argue that these conceptions were not something the children initially possessed. Instead, the context (game of X and O) served as soil for these conceptions to arise.

Our argument is supported by the fact that when no context was given, more than a fifth of the students in our study could not provide an explanation for what ML is. Thus, our findings also challenge the commonly expressed spectacular claims about the contemporary children and youth being an AI-native generation (e.g., DiMaria n.d.; Karampelas, 2023; Ram, 2023; Witt, 2023). It is true that contemporary youth expectedly engage with contemporary technologies more often than their seniors and thus gain more "hands-on" experience at least by being the users of technologies. Based on this study, such arguments belong to the same pool as other unfounded technology-oriented generational dichotomies that claim new generations to possess innate capabilities and knowledge simply because they happened to be born at a certain moment in time (see also Mertala et al., 2024). In contrast, everyday experiences with black box technologies such as ChatGPT may undesirably foster misconceptions; hence, children and youth are at special risk of being exposed to misleading influences in their everyday lives. This further prompts the importance of purposeful, goal-oriented education that provides the students with something they cannot comprehend on their own based on everyday experiences in addition to a conceptual frame within which the newly gained information makes sense.

*5.2. Limitations and suggestions for future research*

One limitation relates to the use of a survey data collection method as it prevents the researcher from asking clarifying questions (like in research interviews). This limitation was partly tackled by expanding the analysis to cover students' responses to the questions on how they would describe what AI and data are. Nevertheless, we encourage future research to apply complementary data collection methods. As our study design was more explorative than explanatory by nature, it restricted us from investigating the possible connections between students' conceptions and their background variables, such as the socioeconomic status of the family, or technological skills, hobbies, or interests of the students. Future research could tackle this issue by collecting larger representative samples with detailed background information.

Furthermore, as discussed in section 4.3, the linguistic context has most likely played a role in the formation of the conception of ML being about learning via or about machines. The Finnish term for machine learning, "koneoppiminen," contains the same word ("kone") as the Finnish word for computer ("tietokone"). The same applies to Swedish ("maskinlärning," "datamaskin") but not to all other languages (cf. "machine learning" and "computer"), and caution is required when making conclusions and generalizations from our findings to other languages. It would be useful to collect comparative data from other countries to identify whether the (mis)conceptions identified in the present study are more universal or influenced by specific languages.

## 5.3. Pedagogical implications

Despite these limitations, our study provides implications for ML- and AI-literacy education, which is as important as ever in the midst of prevalent data-driven systems that have an influence in terms of, for example, privacy, surveillance, profiling, and behavioral engineering (see e.g., Valtonen et al., 2019). First, our findings provide supporting evidence that "language is a powerful tool in shaping thought about abstract domains" (Boroditsky, 2001, p. 1) also in the technological sphere. Therefore, the role of language—words, concepts, metaphors, and the like—should not be overlooked in computer science education. Due to the accumulating evidence about the role of specific linguistic features (see also Mertala, 2019; Mertala & Fagerlund, 2024), it is justified to question whether the universal pedagogical frameworks for (pre)K-12 ML- and AI-literacy education—and computer science education in general—are context (and language) sensitive enough to address more contextual needs.

On another note, several contemporary black box technological platforms that utilize AI, such as voice assistants and LLMs, are presented as agentic beings, in all likelihood contributing to specific kinds of misconceptions via children's everyday experiences (e.g., Szczuka et al., 2022). Previous research has suggested that educators should avoid using language that portrays AI and ML algorithms as sentient and agentic beings to avoid the formation of anthropomorphic misconceptions (Mertala & Fagerlund, 2024). While we agree that this is a valid starting point, it seems not to be enough in itself.

To elaborate further, as previously discussed, different languages provide different connotations for concepts and phenomena: in English, the "computer" is connoted with computing, in Finnish with knowledge, and in Swedish with data, to provide a few examples. Conceptual contextualization, that is, using and teaching concepts in a way that pays attention to the specific educational context—including the linguistic one—(Palsa & Mertala, 2019), might be a useful starting point when going through the "vocabulary of digital technologies" with students.

Secondly, outright teaching students about what such ML techniques as supervised learning and reinforcement learning are should be an obvious pathway to dispel at least the most factually erroneous conceptions of ML, which may often be only superficial (see also Mertala & Fagerlund, 2024). Compelling real-life examples and ways to engage students in a variety of age-appropriate learning activities should be ample, given the popularity of contemporary educational initiatives aiming at bringing AI and ML to schools around the world (e.g., Vartiainen et al., 2021).

**References**

Audry, S. (2021). *Art in the age of machine learning*. MIT Press.

Babari, P., Hielscher, M., Edelsbrunner, P. A., Conti, M., Honegger, B. D., & Marinus, E. (2023). A literature review of children's and youth's conceptions of the internet. *International Journal of Child-Computer Interaction*, 100595. https://doi.org/10.1016/j.ijcci.2023.100595

Barron, A. B., Hebets, E. A., Cleland, T. A., Fitzpatrick, C. L., Hauber, M. E., & Stevens, J. R. (2015). Embracing multiple definitions of learning. *Trends in Neurosciences, 38*(7), 405–407. https://doi.org/10.1016/j.tins.2015.04.008

Biber, Ç., Tuna, A., & Korkmaz, S. (2013). The mistakes and the misconceptions of the eighth grade students on the subject of angles. *European Journal of Science and Mathematics Education*, 1(2), 50–59. https://doi.org/10.30935/scimath/9387

Biesta, G. (2015). What is education for? On good education, teacher judgement, and educational professionalism. *European Journal of Education, 5*0(1), 75–87. https://doi.org/10.1111/ejed.12109

Boeije, H. (2002). A purposeful approach to the constant comparative method in the analysis of qualitative interviews. *Quality and Quantity, 36*, 391–409.https://doi.org/10.1023/A:1020909529486

Boroditsky, L. (2001). Does language shape thought?: Mandarin and English speakers' conceptions of time. *Cognitive Psychology, 43*(1), 1–22. https://doi.org/10.1006/cogp.2001.0748

Davis, K., Christodoulou, J., Seider, S., & Gardner, H. E. (2011). The theory of multiple intelligences. In R.J Sternberg & S.B Kaufman (Eds.), Cambridge Handbook of Intelligence, (pp. 485–503). Cambridge University Press.

De Houwer, J., Barnes-Holmes, D., & Moors, A. (2013). What is learning? On the nature and merits of a functional definition of learning. *Psychonomic Bulletin & Review, 20*, 631–642. https://doi.org/10.3758/s13423-013-0386-3

DiMaria, J. (n.d.) AI-Natives, AI-Immigrants & New Literacies. https://www.digiexam.com/blog/ai-natives-ai-immigrants-new-literacies-digiexam

Druga, S., & Ko, A. J. (2021, June). How do children's perceptions of machine intelligence change when training and coding smart programs? In *Proceedings to Interaction Design and Children (*pp. 49–61). https://doi.org/10.1145/3459990.3460712

Edwards, S., Nolan, A., Henderson, M., Mantilla, A., Plowman, L., & Skouteris, H. (2018). Young children's everyday concepts of the internet: A platform for cyber-safety education in the early years. *British Journal of Educational Technology*, *49*(1), 45–55. https://doi.org/10.1111/bjet.12529

Eskelä-Haapanen, S., & Kiili, C. (2019). 'It Goes Around the World'–Children's Understanding of the Internet. *Nordic Journal of Digital Literacy, 1*4(3–4), 175–187. https://doi.org/10.18261/issn.1891-943x-2019-03-04-0

Fagerlund, J., Leino, K., Kiuru, N., & Niilo-Rämä, M. (2022). Finnish teachers' and students' programming motivation and their role in teaching and learning computational thinking. *Frontiers in Education*, 948783. https://doi.org/10.3389/feduc.2022.948783

Finnish National Agency for Education (2014). Core-curriculum guidelines for basic education. https://www.oph.fi/sites/default/files/documents/perusopetuksen_opetussuunnitelman_perusteet_2014.pdf

Gardner, H., & Hatch, T. (1989). Educational implications of the theory of multiple intelligences. *Educational Researcher, 18*(8), 4–10.

Grönfors, M. (2011). *Laadullisen tutkimuksen kenttätyömenetelmät*. SoFia-Sosiologi-Filosofiapu Vilkka.

Hitron, T., Wald, I., Erel, H., & Zuckerman, O. (2018, June). Introducing children to machine learning concepts through hands-on experience. In *Proceedings of the 17th ACM conference on interaction design and children* (pp. 563–568). https://doi.org/10.1145/3202185.3210776

Jochmann-Mannak, H., Huibers, T., Lentz, L. & Sanders, T. (2010). Children searching information on the Internet: Performance on children's interfaces compared to Google. In: *Towards Accessible Search Systems – Workshop of the 33rd Annual International ACM SIGIR Conference on Research and Development in Information Retrieval*, (pp. 27–35). ACM.

Kammerer, Y. & Bohnacker, M. (2012). Children's web search with Google: the effectiveness of natural language queries. In *Proceedings of the 11th International Conference on Interaction Design and Children*. 12th-15th June, Bremen, Germany, (pp. 184–187).

Karampelas, A. (2023, August). The emergence of AI-natives. https://medium.com/@antonioskarampelas/the-emergence-of-ai-natives-6d67b2543561

Kim, K., Kwon, K., Ottenbreit-Leftwich, A., Bae, H., & Glazewski, K. (2023). Exploring middle school students' common naive conceptions of Artificial Intelligence concepts, and the evolution of these ideas. *Education and Information Technologies*. Advance online publication. https://doi.org/10.1007/s10639-023-11600-3

Kodama, C., St Jean, B., Subramaniam, M., & Taylor, N. G. (2017). There'sa creepy guy on the other end at Google!: engaging middle school students in a drawing activity to elicit their mental models of Google. *Information Retrieval Journal, 20*(5), 403–432. https://doi.org/10.1007/s10791-017-9306-x

Kreinsen, M., & Schulz, S. (2021, October). Students' conceptions of artificial intelligence. In The 16th Workshop in Primary and Secondary Computing Education (pp. 1–2). https://doi.org/10.1145/3481312.3481328

Lachman, S. J. (1997). Learning is a process: Toward an improved definition of learning. *The Journal of Psychology, 131*(5), 477–480. https://doi.org/10.1080/00223989709603535

Legg, S., & Hutter, M. (2007a). A collection of definitions of intelligence. *Arxiv*. https://arxiv.org/pdf/0706.3639.pdf%20a%20collection%20of%20definitions%20of%20intelligence

Legg, S., & Hutter, M. (2007). Universal intelligence: A definition of machine intelligence. *Minds and Machines, 17*, 391–444. https://doi.org/10.1007/s11023-007-9079-x

Mahesh, B. (2020). Machine learning algorithms – a review. International Journal of Science and Research (IJSR), 9(1), 381–386. https://doi.org/10.21275/ART20203995

Marton, F. (1981). Phenomenography—Describing conceptions of the world around us. *Instructional Science*, *10*(2), 177–200. https://doi.org/10.1007/bf00132516

Marx, E., Leonhardt, T., & Bergner, N. (2023). Secondary school students' mental models and attitudes regarding artificial intelligence –A scoping review. *Computers and Education: Artificial Intelligence*, *5*, 100169. https://doi.org/10.1016/j.caeai.2023.100169

Mertala, P. (2019). Young children's conceptions of computers, code, and the Internet. *International Journal of Child-Computer Interaction*, *19*, 56–66. https://doi.org/10.1016/j.ijcci.2018.11.003

Mertala, P. (2020). Young children's perceptions of ubiquitous computing and the Internet of Things. *British Journal of Educational Technology*, *51*(1), 84–102. https://doi.org/10.1111/bjet.12821

Mertala, P. (2022). What, where, when and how: Finnish children's perceptions of learning in preschool. *Early Child Development and Care, 192*(13), 2023–2035. https://doi.org/10.1080/03004430.2021.1970546

Mertala, P., & Fagerlund, J. (2024). Finnish 5th and 6th graders' misconceptions about artificial intelligence. *International Journal of Child-Computer Interaction, 39*, 100630. https://doi.org/10.1016/j.ijcci.2023.100630

Mertala, P., Fagerlund, J., & Calderon, O. (2022). Finnish 5th and 6th grade students' pre-instructional conceptions of artificial intelligence (AI) and their implications for AI literacy education. *Computers and Education: Artificial Intelligence*, *3*, 100095. https://doi.org/10.1016/j.caeai.2022.100095

Mertala, P., Palsa, L., & Slotte Dufva, T. (2020). Monilukutaito koodin purkajana: Ehdotus laaja-alaiseksi ohjelmoinnin pedagogiikaksi. *Media & Viestintä, 43*(1), 21–46. https://doi.org/10.23983/mv.91079

Mertala, P., López-Pernas, S., Vartiainen, H., Saqr, M., & Tedre, M. (2024). Digital natives in the scientific literature: A topic modeling approach. *Computers in Human Behavior, 152*, 108076. https://doi.org/10.1016/j.chb.2023.108076

Mühling, A., & Große-Bölting, G. (2023). Novices' conceptions of machine learning. *Computers and Education: Artificial Intelligence, 4*, 100142.https://doi.org/10.1016/j.caeai.2023.100142

Oinas, S., Vainikainen, M. P., Asikainen, M., Gustavson, N., Halinen, J., Hienonen, N., ... & Hotulainen, R. (2023). *Digitalisaation vaikutus oppimistilanteisiin, oppimiseen ja oppimistuloksiin yläkouluissa: Kansallisen tutkimushankkeen ensituloksia suosituksineen*. Tampere University and the University of Helsinki. https://trepo.tuni.fi/bitstream/handle/10024/145615/978-952-03-2780-4.pdf?sequence=5&isAllowed=y

Ottenbreit-Leftwich, A., Glazewski, K., Jeon, M., Hmelo-Silver, C., Mott, B., Lee, S., & Lester, J. (2021, March). How do Elementary Students Conceptualize Artificial Intelligence? In *Proceedings of the 52nd ACM Technical Symposium on Computer Science Education* (pp. 1261–1261). https://doi.org/10.1145/3408877.3439642

Ottenbreit-Leftwich, A., Glazewski, K., Jeon, M., Jantaraweragul, K., Hmelo-Silver, C. E., Scribner, A., ... & Lester, J. (2022). Lessons Learned for AI Education with Elementary Students and Teachers. *International Journal of Artificial Intelligence in Education*. Advance online publication. https://doi.org/10.1007/s40593-022-00304-3

Ourmazd, A. (2020). Science in the age of machine learning. *Nature Reviews Physics, 2*(7), 342–343.https://doi.org/10.1038/s42254-020-0191-7

Palsa, L., & Mertala, P. (2019). Multiliteracies in local curricula: Conceptual contextualizations of transversal competence in the Finnish curricular framework. *Nordic Journal of Studies in Educational Policy, 5*(2), 114–126. https://doi.org/10.1080/20020317.2019.1635845

Pramling, I. (1988). Developing children's thinking about their own learning. *British Journal of Educational Psychology, 58*(3), 266–278. https://.doi.org/10.1111/j.2044-8279.1988.tb00902.x

Punkari,P. (2022, September). Ursula imuroi ja lataa itsensä – Vatialan koulun oppilaat saavat ohjelmoida robottia. https://yle.fi/a/3-12602471

Ram, P. (2023, December). What the coming age of AI natives portends. https://www.hindustantimes.com/opinion/what-the-coming-age-of-ai-natives-portends-101703936534261.html

Rantala, K. (2019, April). Eskarirobotti opettaa matematiikkaa ja oppii samalla savoa – Vinbotti palkitsee oikean vastauksen rummunpärinällä. https://yle.fi/a/3-10711583

Rubegni, E., Malinverni, L., & Yip, J. (2022, June). "Don't let the robots walk our dogs, but it's ok for them to do our homework": children's perceptions, fears, and hopes in social robots. In *Interaction Design and Children* (pp. 352–361).https://doi.org/10.1145/3501712.3529726

Rücker, M. T., & Pinkwart, N. (2016). Review and discussion of children's conceptions of computers. *Journal of Science Education and Technology*, *25*(2), 274–283. https://doi.org/10.1007/s10956-015-9592-2

Sandberg, A., Broström, S., Johansson, I., Frøkjær, T., Kieferle, C., Seifert, A., … Laan, M. (2017). Children's perspective on learning: An international study in Denmark, Estonia, Germany and Sweden. *Early Childhood Education Journal, 45*(1), 71–81. https://.doi.org/10.1111/j.2044-8279.1988.tb00902.x

Sanusi, I. T., Oyelere, S. S., Vartiainen, H., Suhonen, J., & Tukiainen, M. (2023). A systematic review of teaching and learning machine learning in K-12 education. *Education and Information Technologies, 28*(5), 5967–5997. https://doi.org/10.1007/s10639-022-11416-7

Sarker, I. H. (2021). Machine learning: Algorithms, real-world applications and research directions. *SN Computer Science*, 2(3), 160. https://doi.org/10.1007/s42979-021-00592-x

Selwyn, N., & Gallo Cordoba, B. (2022). Australian public understandings of artificial intelligence. *AI & SOCIETY, 37*(4), 1645¬1662. https://doi.org/10.1007/s00146-021-01268-z

Shamir, G., & Levin, I. (2022). Teaching machine learning in elementary school. *International Journal of Child-Computer Interaction, 31*, 100415. https://doi.org/10.1016/j.ijcci.2021.100415

Sillanpää, M. (2018, December). Robotti aloitti koulunkäynnin Kouvolassa – oppilaat pääsevät kokeilemaan ohjelmointia. https://yle.fi/a/3-10554022

Solyst, J., Axon, A., Stewart, A. E., Eslami, M., & Ogan, A. (2023). *Investigating girls' perspectives and knowledge gaps on ethics and fairness in Artificial Intelligence in a Lightweight workshop*. arXiv preprint arXiv:2302.13947.

Szczuka, J. M., Strathmann, C., Szymczyk, N., Mavrina, L., & Krämer, N. C. (2022). How do children acquire knowledge about voice assistants? A longitudinal field study on children's knowledge about how voice assistants store and process data. *International Journal of Child-Computer Interaction, 33*, 100460. https://doi.org/10.1016/j.ijcci.2022.100460

Tanhua-Piiroinen, E., Kaarakainen, S. S., Kaarakainen, M. T., & Viteli, J. (2020). *Digiajan peruskoulu II*. Opetus- ja kulttuuriministeriön julkaisuja 2020:17. https://julkaisut.valtioneuvosto.fi/bitstream/handle/10024/162236/OKM_2020_17.pdf?sequence=1&isAllowed=y

Taubenfeld, E. (2023, January). 30 funny things you can ask Siri to do. https://www.rd.com/article/funny-things-to-ask-siri/

Thompson, F., & Logue, S. (2006). An exploration of common student misconceptions in science. *International Education Journal*, 7(4), 553–559.

Valtonen, T., Tedre, M., Mäkitalo, K., & Vartiainen, H. (2019). Media Literacy Education in the Age of Machine Learning. *Journal of Media Literacy Education, 1*1(2), 20–36. https://doi.org/10.23860/JMLE-2019-11-2-2

Vandenberg, J., & Mott, B. (2023, June). " AI teaches itself": Exploring young learners' perspectives on Artificial Intelligence for instrument development. *In Proceedings of the 2023 Conference on Innovation and Technology in Computer Science Education* V. 1 (pp. 485–490). https://doi.org/10.1145/3587102.3588778

Vartiainen, H., Toivonen, T., Jormanainen, I., Kahila, J., Tedre, M., & Valtonen, T. (2021). Machine learning for middle schoolers: Learning through data-driven design. *International Journal of Child-Computer Interaction, 29*, 100281. https://doi.org/10.1016/j.ijcci.2021.100281

Vygotsky, L. S. (1987). *The collected works of LS Vygotsky: The fundamentals of defectology* (vol. 2). Springer Science & Business Media.

Wennås Brante, E., & Walldén, R. (2023). "Internet? That's an app you can download". First-graders use linguistic resources to describe internet and digital information. *Education Inquiry*, 14(1), 1–21. https://doi.org/10.1080/20004508.2021.1950273

Witt, T. (2023, April). How AI Natives Will Impact the Workforce. https://accelerationeconomy.com/ai/how-ai-natives-will-impact-the-workforce/#